\documentclass[aps,twocolumn,prl,showpacs,preprintnumbers,amssymb,english]{revtex4-2}
\usepackage{mathrsfs}
\usepackage{hyperref}

\usepackage{microtype}
\usepackage{graphicx}
\usepackage{amsmath, amssymb}
\usepackage{babel}
\usepackage{color}
\usepackage{slashed}
\usepackage{lmodern}
\usepackage[T1]{fontenc}
\usepackage{tikz}
\usepackage{tikz-feynman}
\usepackage{subcaption}
\usepackage{float}
\usepackage{braket}
\usepackage{multirow}

\renewcommand{\thetable}{\Roman{table}}

\def\beqn{\begin{eqnarray}}
\def\eeqn{\end{eqnarray}}
\def\barr{\begin{array}}
\def\earr{\end{array}}
\def\btab{\begin{tabular}}
\def\etab{\end{tabular}}
\def\bite{\begin{itemize}}
\def\eite{\end{itemize}}
\def\bcen{\begin{center}}
\def\ecen{\end{center}}

\newcommand{\sv}[1]{\mathbf{#1}}
\newcommand{\asv}[1]{|\mathbf{#1}|}

\begin{document}

\title{Unique forbidden beta decays at zero momentum transfer}

\author{Chien-Yeah Seng$^{1,2}$}
\author{Ayala Glick-Magid$^{2,3}$}
\author{Vincenzo Cirigliano$^3$}

\affiliation{$^{1}$Facility for Rare Isotope Beams, Michigan State University, East Lansing, MI 48824, USA}
\affiliation{$^{2}$Department of Physics, University of Washington, Seattle, WA 98195-1560, USA}
\affiliation{$^{3}$Institute for Nuclear Theory, University of Washington, Seattle, WA 98195-1550, USA}

\date{\today}

\preprint{NT@UW-24-14}
\preprint{INT-PUB-24-053}

\begin{abstract}
We report an exploratory study of 
the $\mathcal{O}(\alpha)$ structure-dependent electromagnetic radiative corrections to unique first-forbidden nuclear beta decays. We show that the insertion of angular momentum into the nuclear matrix element by the virtual/real photon exchange opens up the decay at vanishing nuclear recoil momentum which was forbidden at tree level, leading to a dramatic change in the decay spectrum not anticipated in existing studies. We discuss its implications for precision tests on the Standard Model and searches for new physics.

\end{abstract}

\maketitle

The standard classification of nuclear beta decays follows 
angular momentum and parity selection rules. Consider
the tree-level amplitude that depends on the nuclear matrix element:
\begin{equation}
	\langle J_f^{P_f}(\sv{p_f})|J^\mu_W(0)|J_i^{P_i}(\sv{p_i})\rangle~,\label{eq:tree}
\end{equation}
where $J_W^\mu$ is the charged weak current, and $J_{i(f)}$, $P_{i(f)}$ and $\sv{p_{i(f)}}$ denote the spin, parity, and momentum of the initial (final) nucleus, and we define the nuclear recoil momentum $\sv{q}\equiv \sv{p_i}-\sv{p_f}$. Decays that satisfy the  selection rules: $|J_i-J_f|=0,1$ and $P_i=P_f$ are known as ``allowed'' beta decays, while the rest are known as ``forbidden'' decays as their tree-level transition matrix elements are kinematically suppressed. In particular, for decays with $|J_i-J_f|\geq 2$, the amplitude above survives only when $\sv q$ is non-zero, due to rotational invariance. 

Allowed beta decays provide stringent tests of the Standard Model (SM) and probe physics beyond the Standard Model (BSM)~\cite{Gorchtein:2023naa}, e.g. through the precise measurement of the Cabibbo-Kobayashi-Maskawa (CKM) matrix elements~\cite{Cabibbo:1963yz,Kobayashi:1973fv}, and by constraining exotic interactions~\cite{Cirigliano:2012ab,Cirigliano:2013xha,Gonzalez-Alonso:2018omy,Cirigliano:2023nol,johnson1963precision,gluck1998order,glick2022nuclear,muller2022beta,Mishnayot:2021uki,longfellow2024improved}.
On the other hand,
forbidden decays have recently received increased attention due to their complementary role in probing new physics~\cite{Brodeur:2023eul}, e.g. exotic  (non $V$-$A$) charged-current couplings 
(see Refs.~\cite{Cirigliano:2012ab,Gonzalez-Alonso:2018omy} for a mapping between the traditional Lee-Yang~\cite{Lee:1956qn}
nucleon-level interaction and the modern Standard Model Effective Field theory framework).
It has recently been highlighted that measurements of the unique forbidden beta decay spectrum provide simultaneous access to both the Fierz term and the electron-neutrino angular correlation~\cite{Glick-Magid:2016rsv, glick2023multipole};
the former is linear in the coefficients of new physics but lacks sensitivity to right-handed neutrino interactions, while the latter is sensitive to both left- and right-handed neutrino interactions but is quadratic in the new physics coefficients.
As an example, we consider the unique first forbidden decay ($|J_i-J_f|=2$, $P_i=-P_f$), whose leading order matrix element is suppressed linearly by the nuclear momentum transfer $\asv{q}$. 
The tree-level differential  decay rate takes the form
\begin{multline}
  \label{eq:spect2-}
  \frac{d\Gamma_{\text{tree}}}{dE_ed\Omega} \propto 
   \asv q^2 \left\{1 +b\,\frac{m_{e}}{E_e}
  +a\left[2\boldsymbol{\beta}\cdot\sv{\hat{p}_\nu}
-\sv{\hat{p}_\nu}\cdot\sv{\hat{q}}\:\boldsymbol{\beta}\cdot\sv{\hat{q}}\right]\right\}
    \text{,}
\end{multline}
where $\asv q^2$ comes from the nuclear matrix element squared, $\sv{\hat{p}_\nu}$ is the unit neutrino momentum, $m_e$, $E_e$, $\sv{p_e}$ are the mass, energy, momentum of the emitted electron, and $\boldsymbol{\beta}\equiv\sv{p_e}/E_e$. 
The observables depending on the BSM tensor coefficients 
$C_T^{\left(\prime \right)}$~\cite{Lee:1956qn,Jackson:1957auh,Jackson:1957zz}  
are the Fierz term 
$b=\pm \mathfrak{Re} [(C_{T}+C_{T}^{'})/C_{A} ] $,
and  the angular correlation 
$a=- (1/5) (1- (|C_{T}|^{2}+|C_{T}^{'}|^{2})/|C_{A}|^{2})$.
The last term multiplied by $a$ in the tree-level rate does not exist in allowed decays. This term prevents the angular correlation from vanishing as in the allowed decays when integrating over the angles, making the unique forbidden spectrum sensitive to $a$, and as a result, also to right-handed tensor couplings.

This observation has motivated a number of new experiments to study unique first-forbidden decays. Measurements of ${}^{90}$Sr$(0^+)\to {}^{90}$Y$(2^-)$ and ${}^{90}$Y$(2^-)\to {}^{90}$Zr$(0^+)$ are currently being conducted at the Hebrew University of Jerusalem, and these will be followed with measurements of ${}^{16}$N$(2^-)\to {}^{16}$O$(0^+)$, with an aim for $10^{-3}$ accuracy ~\cite{mardor2018soreq, ohayon2018weak, ron2024private}. Additionally, studies on ${}^{90}$Y$(2^-)\to {}^{90}$Zr$(0^+)$ and ${}^{144}$Pr$(0^-)\to {}^{144}$Nd$(2^+)$ are underway at the Oak Ridge National Laboratory, aiming for 1-2\% accuracy at the first stage~\cite{rasco2024private, shuai2022determination}.
However, similar to their allowed counterparts, one requires all the SM predictions of the forbidden decays to reach the same accuracy in order to maximize the discovery potential of the experiments. Existing theory analyses of forbidden beta decays focus mainly on calculations of tree-level transition amplitudes, Coulomb effects, shape factor, recoil corrections~\cite{Weidenmuller:1961zz,Damgaard:1969zz,Bertsch:1970zz,Smith:1970zjs,Vergados:1971wjf,VanEijk:1971zkr,Schweitzer:1972ynq,Smith:1972zza,Schweitzer:1973ccr,Smith:1973yru,Lakshminarayana:1981zz,Becker:1984kdt,Civitarese:1986wgx,Warburton:1990zz,Warburton:1991zzd,Warburton:1992zz,Suhonen:1993jwx,Martinez-Pinedo:1998haz,Borzov:2003bb,Mustonen:2006qn,Haaranen:2014rga,Fang:2015cma,Nabi:2016nnv,Haaranen:2016rzs,Nabi:2017zxy,Kostensalo:2017xxq,Kumar:2020vxr,Kumar:2021ffu,Glick-Magid:2021xty,Sharma:2022kxf,Sharma:2022aws,Wang:2023jil,Kumar:2024isb,Saxena:2024muu,DeGregorio:2024ivh} and structure-independent, ``outer'' radiative corrections (RC)~\cite{wilkinson1998evaluation}. Existing phenomenological calculations of tree-level decay matrix elements of heavy nuclei (e.g. $A\sim 90$) typically have uncertainties spanning an order of magnitude~\cite{Vergados:1971wjf, Civitarese:1986wgx, Nabi:2017zxy}, which will be improved with future \textit{ab initio} calculations, e.g.~\cite{glick2024ab}.
However, an important missing piece in the program is the study of the full one-loop, structure-dependent radiative correction to the forbidden decay amplitude; the latter is known to play a central role in the interpretation of precision beta decays, e.g. the extraction of $V_{ud}$~\cite{Seng:2018yzq,Seng:2018qru,Shiells:2020fqp,Seng:2021syx,Gorchtein:2023srs,Cirigliano:2023fnz,Cirigliano:2024msg,Cirigliano:2024rfk,Gennari:2024sbn}, the nucleon axial coupling constant~\cite{Hayen:2020cxh,Gorchtein:2021fce,Cirigliano:2022hob,Seng:2024ker}, and the correction to the beta spectrum~\cite{Hill:2023acw,Hill:2023bfh,Borah:2024ghn}.

In this Letter, we report the first study of the $\mathcal{O}(\alpha)$ RC to forbidden decays which leads to an interesting new observation: The 
usual statement
that forbidden decay amplitudes with $|J_i-J_f|\geq 2$ vanish in the non-recoil limit is falsified by the RC due to the introduction of an extra current operator into the nuclear matrix element that alleviates the inhibition from the angular momentum difference. 
As a consequence,
at small enough $\asv q$ the RC amplitude actually overtakes Eq.\eqref{eq:spect2-} as the main contributor to the forbidden decay rate. The same effect can also be achieved with new light degrees of freedom (DOFs) in the BSM sector that take the role of the photon in the RC.
Explicitly, the differential decay rate now takes the form:
\begin{equation}
    d\Gamma\propto f_0 \asv q^0+f_1\asv q^1+f_2\asv q^2~, 
\end{equation}
where the first two terms probe the SM RC and light new physics, while the third term probes the SM tree-level effects and heavy new physics. 
Therefore, the precise study of the $\asv q\rightarrow 0$ behavior of forbidden decays provides a unique opportunity to simultaneously probe higher-order SM physics as well as BSM physics, without being contaminated by the large SM tree-level uncertainty~\footnote{There exists exploratory studies of radiation-induced selection-rule violations, e.g. Refs.\cite{Longmire:1949zz,merzbacher1951note,hanawa1952radiative} which were done prior to the establishment of the standard $V-A$ theory and are not straightforwardly applicable to modern analyses of beta decays. Also, they focused on corrections to the total lifetime instead of the $q$-dependence of the decay rate. Some other works, e.g. Refs.\cite{yem1966theoretical,Ford:1969gzz,Draxler:1978bhv,rose1962beta,Pfutzner:2015nra,Pacheco:1987yq}, focused on the spectrum distortion due to bremsstrahlung instead of one-loop diagrams as in this work.}. We investigate this novel idea in detail and discuss future prospects. 

\begin{figure}[htb]
	\centering
	\includegraphics[width=0.45\columnwidth]{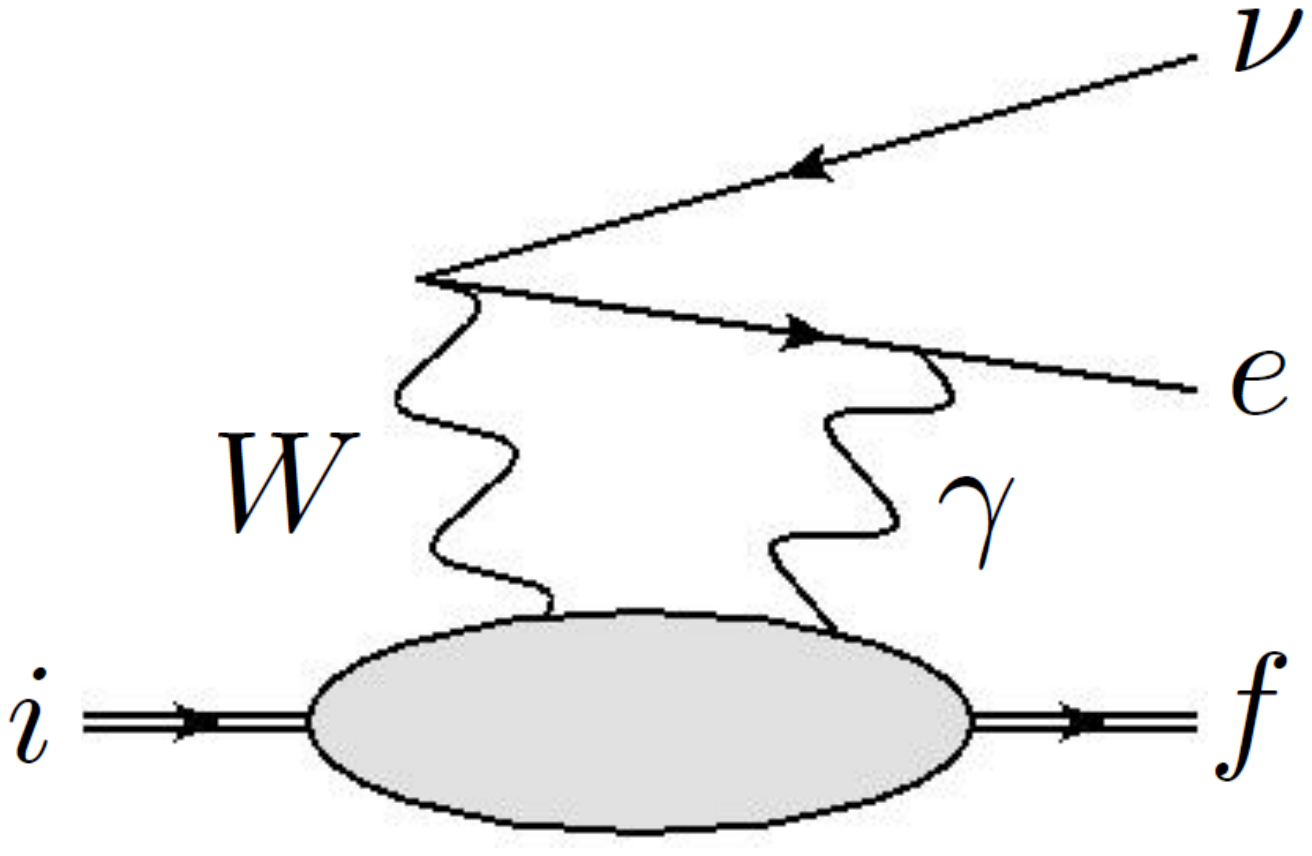}
	\includegraphics[width=0.45\columnwidth]{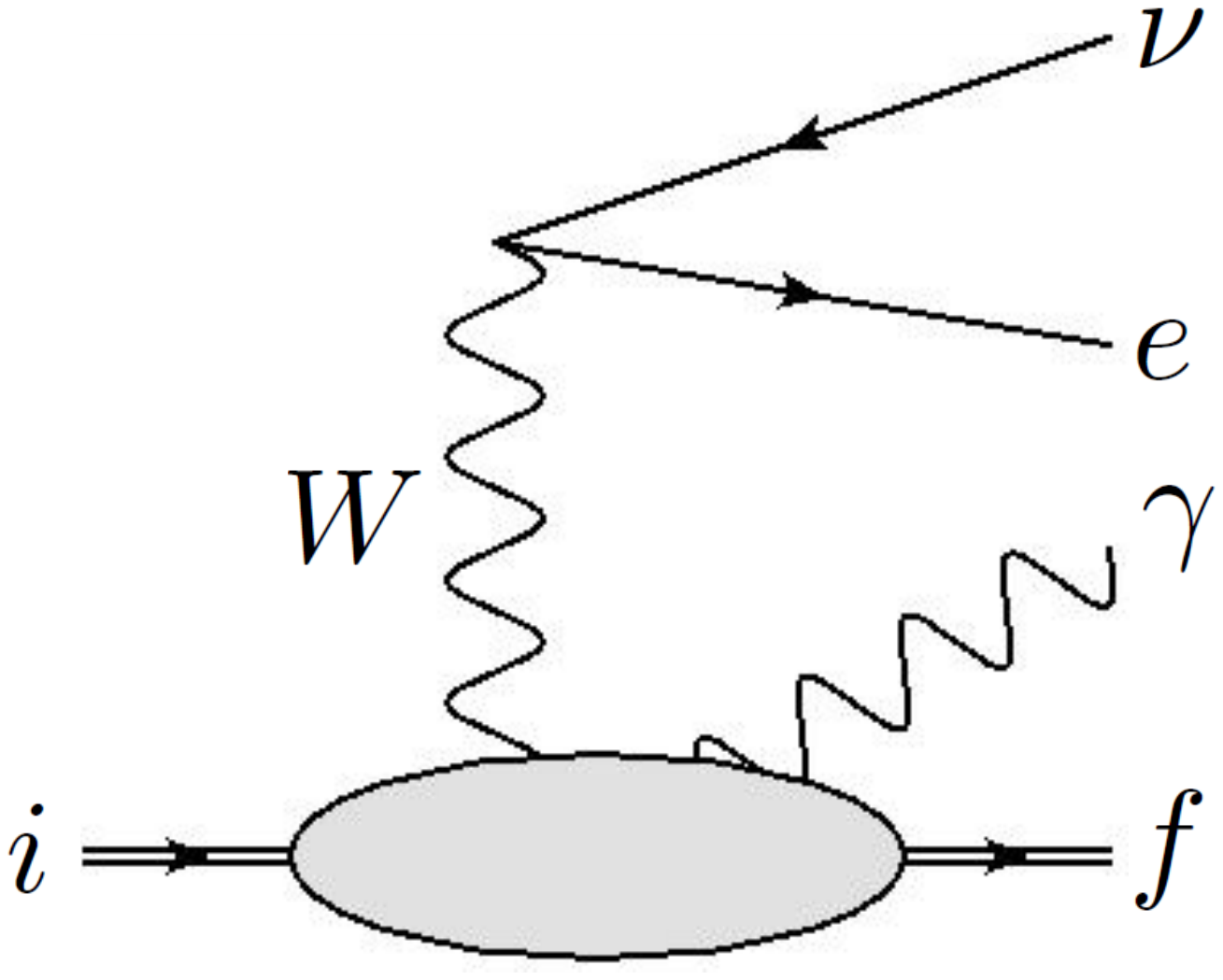}
	\caption{\label{fig:RC}$\mathcal{O}(\alpha)$ Feynman diagrams that open up the forbidden nuclear transition at $\asv q=0$. } 
\end{figure}

We begin by studying the SM RC. It can be recognized that among all the $\mathcal{O}(\alpha)$ corrections, only the two diagrams in Fig.\ref{fig:RC}, namely the $\gamma W$-box diagram and the bremsstrahlung diagram with a photon emitted by the nucleus, can lead to a non-zero amplitude at $\asv q=0$, since all other diagrams depend on the tree-level nuclear matrix element in Eq.\eqref{eq:tree} (although $J_W^\mu$ may be renormalized) which has to satisfy the same angular momentum and parity selection rules. Their corresponding amplitudes read (assuming $\beta^-$-decay)~\cite{Seng:2021syx}:
\begin{eqnarray}
		&&\mathcal{M}_{\gamma W}(p_e)  =  \frac{G_{F}V_{ud}}{\sqrt{2}}e^{2}L_{\lambda}\int\frac{d^{4}k}{(2\pi)^{4}}\frac{M_{W}^{2}}{M_{W}^{2}-k^{2}}\frac{1}{k^{2}-m_{\gamma}^{2}+i\varepsilon}\nonumber \\
		&  & \times\frac{1}{(p_{e}-k)^{2}-m_{e}^{2}+i\varepsilon}\{ -2g^{\nu\lambda}p_{e}^{\mu}+g^{\mu\lambda}k^{\nu}+g^{\nu\lambda}k^{\mu}\nonumber\\
		&&-g^{\mu\nu}k^{\lambda}-i\epsilon^{\mu\nu\alpha\lambda}k_{\alpha}\} T_{\mu\nu}(k)~,\nonumber\\
		&&\mathcal{M}_{\text{brem}}(p_e)=i\frac{G_{F}V_{ud}}{\sqrt{2}}e\epsilon^{\mu*}L^{\nu}T_{\mu\nu}(k)~,
\end{eqnarray}
where $L_\lambda=\bar{u}_e\gamma_\lambda(1-\gamma_5)v_\nu$ is the lepton current, and 
\begin{equation}
	T^{\mu\nu}(k)\equiv\int d^{4}xe^{ik\cdot x}\langle\phi_{f}(-\sv{q})|T[J_\text{em}^{\mu}(x)J_{W}^{\nu}(0)]|\phi_{i}(\sv{0})\rangle
\end{equation}
is a ``generalized Compton tensor'' involving the electromagnetic (${\rm em}$) and weak  ($W$) currents, with external momenta  $\sv{p_i}= \sv{0}$ and $\sv{p_f}=-\sv{q}$.  
We focus on the $\gamma W$-box diagram that gives the dominant contribution as we show later.

To be concrete, let us concentrate on unique first-forbidden decays involving the transition $2^-\leftrightarrow 0^+$, in accordance with the planned experiments we mentioned in the introduction. The first important observation is that at  $\asv q=0$ the loop integral in $\mathcal{M}_{\gamma W}$ is dominated by  small values of the virtual photon momentum $k$. This is seen by noticing that when $k$ is large, one may take $p_e\rightarrow 0$ in the integrand which reduces the integral to:
\begin{equation}
	\mathcal{M}_{\gamma W}\rightarrow \frac{G_FV_{ud}}{\sqrt{2}}e^2L_\lambda \langle \phi_f(\sv{0})|K^\lambda|\phi_i(\sv{0})\rangle~,
\end{equation}
where 
\begin{eqnarray}
&&K^\lambda\equiv \int\frac{d^4k}{(2\pi)^4}\frac{M_W^2}{M_W^2-k	^2}\frac{1}{(k^2)^2}\{g^{\mu\lambda}k^\nu+g^{\nu\lambda}k^\mu\nonumber\\
&&-g^{\mu\nu}k^\lambda-i\epsilon^{\mu\nu\alpha\lambda}k_\alpha\}\int d^4xe^{ik\cdot x}T[J^\text{em}_\mu(x)J^W_\nu(0)]~.
\end{eqnarray}
No matter how complicated $K^\lambda$ is, it remains an ordinary four-vector with no external momentum dependence, so $\langle\phi_f(\sv{0})|K^\lambda|\phi_i(\sv{0})\rangle$ must vanish due to rotational invariance given that $|J_i-J_f|=2$. 
Hence,   the integral is dominated by the small-$k$ region, 
and more precisely the  ultrasoft  photon region 
in which $k^0 \sim |\sv{k}| \sim E_e$ (see Refs.~\cite{Cirigliano:2024msg,Cirigliano:2024rfk} for a discussion of radiative corrections to superallowed $\beta$ decays in terms of 
various regions in photon virtuality).
In the ultrasoft region $T_{\mu \nu} (k)$ becomes sensitive to the 
nuclear states $|X\rangle$ that can be reached from the initial and final states through 
insertions of the weak and electromagnetic currents, and can be written as:
\begin{eqnarray}
	&&T^{\mu\nu}(k)  \approx -i\sqrt{4M_{i}M_{f}}\sum_{X}\sum_{m_{X}}\nonumber\\
	&&\left\{ \frac{\langle J_{f}m_{f}|J_{\text{em}}^{\mu}(\sv{k})|J_{X}m_{X}\rangle\langle J_{X}m_{X}|J_{W}^{\nu}(-\sv{k})|J_{i}m_{i}\rangle}{M_{X}-(M_{f}+k_{0}+i\varepsilon)}\right.\nonumber \\
	&  & \left.+\frac{\langle J_{f}m_{f}|J_{W}^{\nu}(-\sv{k})|J_{X}m_{X}\rangle\langle J_{X}m_{X}|J_{\text{em}}^{\mu}(\sv{k})|J_{i}m_{i}\rangle}{M_{X}-(M_{i}-k_{0}+i\varepsilon)}\right\}~,
\label{eq:tmunu}
\end{eqnarray}
where we take $\sv{q}=0$
and all states are normalized to 1.

In the small-$k$ region, we can take $\asv{k}R$ as a small expansion parameter, where $R$ is a nuclear radius. This allows us to apply the standard multipole expansion of the Fourier-transformed current operators $J^\mu(\pm \sv{k})$~\cite{walecka2004theoretical}:
\begin{eqnarray}
	&&J^0(\pm\sv{k})=\sqrt{4\pi}\sum_{J=0}^{\infty}(\mp i)^J[J]C_{J0}(\asv{k})~,\nonumber\\
	&&\sv{J}(\pm\sv{k})=\pm\sqrt{4\pi}\sum_{J=0}^\infty(\mp i)^J[J]L_{J0}(\asv{k})\boldsymbol{\epsilon_0}^*~,\nonumber\\
	&&-\sqrt{2\pi}\sum_{\lambda=\pm 1}\sum_{J=1}^\infty(\mp i)^J[J](\lambda M_{J\lambda}(\asv{k})\mp E_{J\lambda}(\asv{k}))\boldsymbol{\epsilon_\lambda}^*~,
\end{eqnarray}
where $[J]=\sqrt{2J+1}$; here we introduce $C$, $L$, $M$ and $E$ as the Coulomb, longitudinal, transverse magnetic and transverse electric multipole operators respectively, with the polarization vectors $\boldsymbol{\epsilon_0}=\sv{\hat{z}}$, $\boldsymbol{\epsilon_{\pm 1}}=\mp (\sv{\hat{x}}\pm i\sv{\hat{y}})/\sqrt{2}$ defined in a coordinate frame with $\sv{\hat{z}}\equiv\sv{\hat{k}}$. Following the power counting in the multipole formalism~\cite{Glick-Magid:2021xty}, we find that the leading contributors to $T^{\mu\nu}$ for the $i(2^-_g)\to f(0^+_g)$ 
transition (here $g$ stands for ground state)  involve  the ground states $i(2_g^-),f(0_g^+)$ 
and the $J=1$ excited states $i(1_X^+),f(1_X^-)$. 
While the full leading expression of $T^{\mu\nu}$ can be found in the supplementary material, we observe that the ground state contribution involves the electromagnetic Coulomb operator and is enhanced by the atomic number $Z$. It gives rise to:
\begin{eqnarray}
	&&T^{0j}(m)  \approx  i\sqrt{16\pi M_{i}M_{f}}\asv{k}\mathfrak C_{\text{tree}}\left(\frac{Z_{f}\mathfrak C_{\gamma g}^f}{k_{0}+i\varepsilon}-\frac{Z_{i}\mathfrak C_{\gamma g}^i}{k_{0}-i\varepsilon}\right)\times\nonumber \\
	&  & \left\{ S_{0m}(\theta)(\boldsymbol{\epsilon_{0}}^{*})^{j}\mp\frac{\sqrt{3}}{2}\left(S_{\mp1m}(\theta)(\boldsymbol{\epsilon_{1}}^{*})^{j}+S_{\pm1m}(\theta)(\boldsymbol{\epsilon_{-1}}^{*})^{j}\right)\right\} \nonumber \\\label{eq:T0ibig}
\end{eqnarray}
where the upper (lower) sign corresponds to the $i(2^{-})\to f(0^{+})$
($i(0^{+})\to f(2^{-})$) decay, $\{\mathfrak C_\text{tree},\mathfrak C_{\gamma g}^{i,f}\}$ are reduced nuclear matrix elements (non-zero at $\asv{k}=0$) defined in Table~\ref{tab:RME} in the supplementary material, and $m$ is the magnetic
quantum number of the external $2^{-}$ nuclear state along $\sv{\hat{p}_e}$. The matrix $S(\theta)$ (where $\theta=\cos^{-1}(\sv{\hat{p}_e}\cdot\sv{\hat{k}})$), whose explicit expression can be found in the supplementary material, rotates the third axis of the $2^-$ state's spin from $\sv{\hat{p}_e}$ to $\sv{\hat{k}}$; the latter is needed for the proper application of the Wigner-Eckart theorem involving the multipole operators. 

We may now evaluate the box diagram amplitude $\mathcal{M}_{\gamma W}$. First, to suppress the dependence on physics at large $k$, we make use of our previous argument that the  amplitude vanishes at $p_e\rightarrow 0$ to 
write the amplitude in the subtracted form
\begin{eqnarray}
	\mathcal{M}_{\gamma W}(p_e)&=& \mathcal{M}_{\gamma W}(p_e)-\mathcal{M}_{\gamma W}(0)~.
\end{eqnarray}
We then
substitute the leading small-$k$ expression of $T^{\mu\nu}$ given in Eq.\eqref{eq:tmunu} 
in the integrand appearing in both $\mathcal{M}_{\gamma W}(p_e)$ and $\mathcal{M}_{\gamma W}(0)$.
We may then evaluate the $k_0$-integral by closing up the contour from the lower half in the complex $k_0$-plane (which is an arbitrary choice; one may also choose the upper half). In doing so we observe that, only the residue at $k_0=-i\varepsilon$ is enhanced by the atomic number $Z_f$ at small $\asv{k}$; picking up other poles always leads to a partial cancellation between the two elastic terms in Eq.\eqref{eq:T0ibig}, $Z_f/k_0-Z_i/k_0=1/k_0$, that loses such an enhancement. It is also easy to see that the bremsstrahlung amplitude $\mathcal{M}_\text{brem}$ does not receive such an enhancement, because there the $\pm i\varepsilon$ in $T^{\mu\nu}$ does not play a role and the partial cancellation always takes place. Retaining only the $Z_f$-enhanced term~\footnote{Without an explicit calculation, one cannot exclude that the sum over the excited intermediate 
state may make up for the $Z_f$ factor. However,  even if this were the case, it is unlikely that such neglected terms exactly cancel the contribution we focus on here, absent a symmetry to enforce the cancellation. 
Moreover, in radiative corrections to other systems~\cite{Gennari:2024sbn}, no strong sensitivity to highly excited nuclear states is observed.} leads to, after straightforward algebra:
\begin{equation}
	\mathcal{M}_{\gamma W}(p_e)\approx -\frac{G_FV_{ud}}{\sqrt{2}}e^2\left(L^0I^0-\sv{L}\cdot\sv{\hat{p}_e}I'\right)~,\label{eq:MgammaW}
\end{equation}
where we are left with two scalar integrals:
\begin{eqnarray}
	&&I^0=\frac{Z_f\sqrt{16\pi M_iM_f}}{2\pi^2}|\sv{p_e}|\int_0^\pi d\theta\sin\theta\cos\theta S_{0m}(\theta)\nonumber\\
	&&\times\int_{0}^{\infty}d\asv{k}\frac{\mathfrak C_\text{tree}(\asv{k})\mathfrak C_{\gamma g}^f(\asv{k})}{\asv{k}-2|\sv{p_e}|\cos\theta}~,\nonumber\\
	&&I'=\frac{Z_f\sqrt{16\pi M_iM_f}}{4\pi^2}\int_0^\pi d\theta\sin\theta\int_{0}^{\infty}d\asv{k}\frac{\mathfrak C_\text{tree}(\asv{k})\mathfrak C_{\gamma g}^f(\asv{k})}{\asv{k}-2|\sv{p_e}|\cos\theta} \nonumber\\
	&&\times\left\{2E_e\cos\theta S_{0m}(\theta)\mp\sqrt{\frac{3}{2}}S_{\mp 1m}(\theta)\sin\theta(|\sv{p_e}|\cos\theta-E_e)\right.\nonumber\\
	&&\left.\mp\sqrt{\frac{3}{2}}S_{\pm 1m}(\theta)\sin\theta(|\sv{p_e}|\cos\theta+E_e)\right\}~.\label{eq:I0Iprime}
\end{eqnarray}

The above integrals are logarithmically divergent in the ultraviolet. In an effective field theory (EFT) approach, 
one would regulate the integrals in dimensional regularization and reabsorb the divergence through terms from the potential photon region~\cite{Cirigliano:2024msg}. 
Here, however, we are interested in a first 
rough estimate and therefore introduce the 
$\asv{k}$-dependence of $\mathfrak C_{\gamma g}^f$ and $\mathfrak C_\text{tree}$ 
as a means
 to ensure the ultraviolet-finiteness of the $\asv{k}$-integral and obtain a model-dependent value for the corresponding EFT coupling.
In principle, $\mathfrak C_{\gamma g}^f$
can be inferred from the nuclear charge distribution data and the latter requires nuclear structure calculations, but in this Letter we resort to a simple approximation for illustration. First, we know the small-$\asv{k}$ expansion of the charge form factor: $\mathfrak C_{\gamma  g}(\asv{k})=1-\asv{k}^2R_C^2/6+\dots$, 
where $R_C$ is the nuclear root-mean-square charge radius. So, we adopt a simple monopole expression for the charge form factor (with $\Lambda^2=6/R_C^2$) that reproduces the leading term in the small-$\asv{k}$-expansion. We assume the same form factor in $\mathfrak C_\text{tree} $ for simplicity:
\begin{equation}
	\mathfrak C_{\gamma g}(\asv{k})\approx \frac{\Lambda^2}{\Lambda^2+\asv{k}^2}~,~\mathfrak C_\text{tree}(\asv{k})\approx \mathfrak C_\text{0}^A\frac{\Lambda^2}{\Lambda^2+\asv{k}^2}~,
\end{equation} 
since no extra information of the latter is currently available. With these, the integrals $I^0$ and $I'$ can be evaluated analytically, and the squared amplitude $|\mathcal{M}_{\gamma W}|^2$ as well as the tree-loop interference $2|\mathcal{M}_\text{tree}^*\mathcal{M}_{\gamma W}|$, after averaging and summing over initial and final spins, are found to be:
\begin{eqnarray}
	&&|\overline{\mathcal{M}_{\gamma W}}|^2\approx\frac{4M_iM_f}{2J_i+1}\frac{3\pi}{2}G_F^2V_{ud}^2(Z_f\alpha)^2|\mathfrak C_0^A|^2E_eE_\nu\nonumber\\&&\times\left\{8E_e^2L_\Lambda^2-\frac{|\sv{p_e}|^2}{32}\left(4L_\Lambda-3\right)\left(44L_\Lambda+15\right)\right.\nonumber\\
	&&\left.+\boldsymbol{\beta}\cdot\sv{\hat{p}_\nu}\left[6E_e^2L_\Lambda+\frac{5|\sv{p_e}|^2}{32}\left(4L_\Lambda-3\right)^2\right]\right\}~,\nonumber\\
    &&2\mathfrak
    {Re}\{\overline{\mathcal{M}_\text{tree}^*\mathcal{M}_{\gamma W}}\}\approx -\frac{4M_iM_f}{2J_i+1}\frac{3\pi}{2}G_F^2V_{ud}^2Z_f\alpha|\mathfrak C_0^A|^2\asv q\nonumber\\&&\times\sin\theta_{eq}\left\{16E_e^2E_\nu L_\Lambda-2E_e|\sv{p_e}|^2\left(4L_\Lambda+3\right)\right.\nonumber\\
    &&\left.+E_e|\sv{p_e}|\asv q\cos\theta_{eq}\left(20L_\Lambda+9\right)-2E_\nu|\sv{p_e}|^2\left(4L_\Lambda-3\right)\right\}~,\label{eq:central}
\end{eqnarray}
where $L_\Lambda\equiv\ln(\Lambda/|\sv{p_e}|)$~\footnote{Using dimensional regularization one obtains the same result with the replacement $\ln\Lambda \to \ln \mu + {\rm constant}$}.
The appearance of the angle $\theta_{eq}\equiv\cos^{-1}(\sv{\hat{p}_e}\cdot\sv{\hat{q}})$ in the second expression is because one needs to re-align the nuclear polarization direction in $\mathcal{M}_\text{tree}$ from $\sv{\hat{q}}$ to $\sv{\hat{p}_e}$ in order to interfere with $\mathcal{M}_{\gamma W}$ in Eq.\eqref{eq:MgammaW}. 
Recall that the result above is derived by taking $\sv{q}\approx\sv{0}$ in $T^{\mu\nu}$, and the non-vanishing  of $|\overline{\mathcal{M}_{\gamma W}}|^2$ demonstrates our assertion at the beginning of this Letter. Notice, however, that one still keeps the finite $\sv{q}$ in the momentum conservation $\sv{q}=\sv{p_e}+\sv{p_\nu}$, and Eq.\eqref{eq:central} may still be applicable for small but non-zero values of $\asv q$. 

\begin{figure}[htb]
	\centering
	\includegraphics[width=0.9\columnwidth]{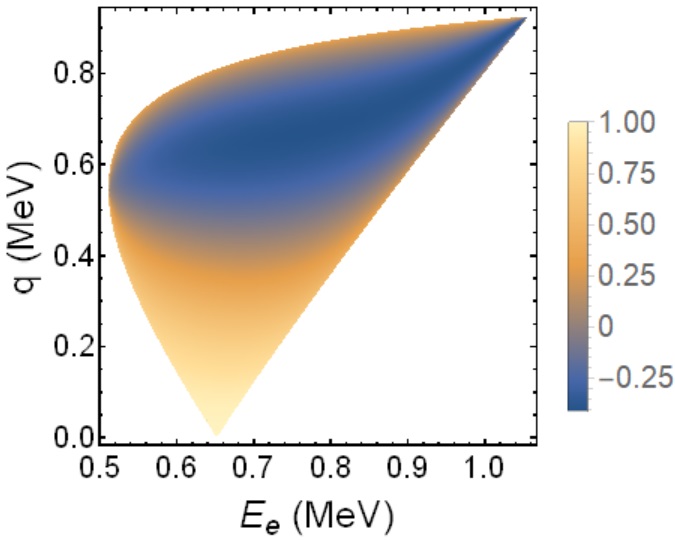}
	\caption{\label{fig:SrtoY}The Dalitz plot of $1-|\overline{\mathcal{M}_\text{tree}}|^2/|\overline{\mathcal{M}_\text{tot}}|^2$ for $^{90}$Sr$\to{}^{90}$Y.}
\end{figure}

\begin{figure}[htb]
	\centering
	\includegraphics[width=0.9\columnwidth]{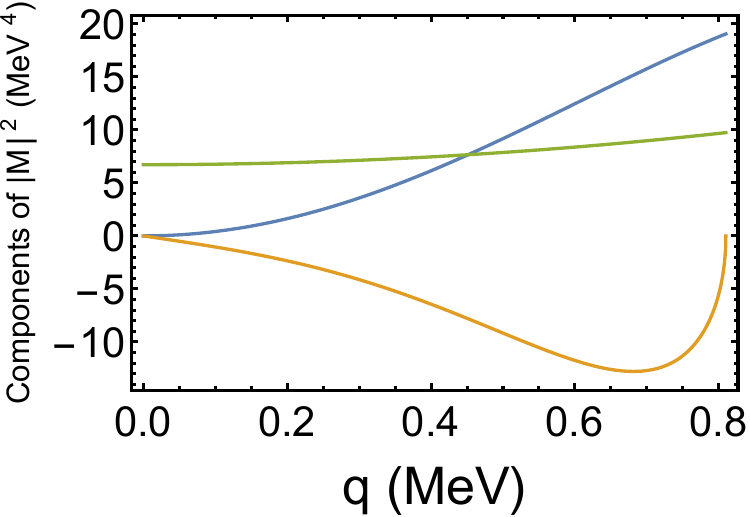}
	\caption{\label{fig:SrtoYfixEe}The plot of $|\overline{\mathcal{M}_\text{tree}}|^2$ (blue), $2\mathfrak
    {Re}\{\overline{\mathcal{M}_\text{tree}^*\mathcal{M}_{\gamma W}}\}$ (orange) and $|\overline{\mathcal{M}_{\gamma W}}|^2$ (green) at fixed $E_e$ for $^{90}$Sr$\to{}^{90}$Y, scaling out the constant $4M_iM_fG_F^2V_{ud}^2|\mathfrak C_0^A|^2/(2J_i+1)$.}
\end{figure}

It is instructive to compare the radiative terms to the tree-level squared amplitude~\cite{Glick-Magid:2016rsv}:
\begin{eqnarray}
	&&|\overline{\mathcal{M}_\text{tree}}|^2\approx \frac{4M_iM_f}{2J_i+1}16\pi G_F^2V_{ud}^2|\mathfrak C_0^A|^2E_eE_\nu \asv q^2\nonumber\\
	&&\times\left(\frac{5}{2}-\boldsymbol{\beta}\cdot\sv{\hat{p}_\nu}+\frac{\sv{\hat{p}_\nu}\cdot\sv{\hat{q}}\:\boldsymbol{\beta}\cdot\sv{\hat{q}}}{2}\right)~.\label{eq:Mtree2}
\end{eqnarray}
We do this for the transition $^{90}$Sr$\to{}^{90}$Y, which is interesting due to its large $Z_f$ and a particularly small $Q_\beta$-value of 545.9(14) keV; $R_C\approx 4.26$~fm is taken for the nuclear radius~\cite{Angeli:2013epw}. First, we plot the quantity $1-|\overline{\mathcal{M}_\text{tree}}|^2/|\overline{\mathcal{M}_\text{tot}}|^2$ in the full 3-body phase space ($\mathcal{D}_3$) spanned by $E_e$ and $\asv q$:
\begin{eqnarray}
    C(\asv q)-D(\asv q)&<E_e<&C(\asv q)+D(\asv q)~,\nonumber\\
    0&<\asv q<&\sqrt{E_{f,\text{max}}^2-M_f^2}
\end{eqnarray}
where 
\begin{eqnarray}
    C(\asv q)&=&\frac{(M_i-E_f)(M_i^2+m_e^2+M_f^2-2E_f M_i)}{2(M_i^2+M_f^2-2E_fM_i)}\nonumber\\
    D(\asv q)&=&\frac{\asv q(M_i^2+M_f^2-m_e^2-2E_fM_i)}{2(M_i^2+M_f^2-2E_fM_i)}\nonumber\\
    E_{f,\text{max}}&=&\frac{M_i^2-m_e^2+M_f^2}{2M_i}~,
\end{eqnarray}
with $E_f\equiv\sqrt{M_f^2+\asv q^2}$, and $|\overline{\mathcal{M}_\text{tot}}|^2$ sums the three terms in Eqs.\eqref{eq:central} and \eqref{eq:Mtree2}. From Fig.\ref{fig:SrtoY} it is clearly seen that the size of the radiative corrections is substantial, and overtakes the tree-level contribution in the small-$q$ region which constitutes a significant portion of the entire phase space. 

In Fig.\ref{fig:SrtoYfixEe} we plot the various terms in Eqs.\eqref{eq:central}, \eqref{eq:Mtree2}on a fixed-energy slice in the phase space: $E_e=((M_i-M_f)^2+m_e^2)/(2(M_i-M_f))$, that includes the $\asv q=0$ point. One sees that, at $\asv q\rightarrow 0$ the tree-level squared amplitude and the interference term decay as $\asv q^2$ and $\asv q$ respectively, while $|\overline{\mathcal{M}_{\gamma W}}|^2$ approaches a constant, which leads to its dominance at the small-$\asv q$ region, in stark contrast to the traditional understanding of forbidden decays; therefore, any precision study of the decay shape without including the RC would be premature. We can also study the effect of RC on the total decay rate using the formula:
\begin{equation}
    \Gamma=\frac{1}{64\pi^3M_i}\int_{\mathcal{D}_3}dE_e d\asv q\frac{\asv q}{E_f}|\mathcal{M}|^2~.\label{eq:totalrate}
\end{equation}
For $^{90}$Sr$\to{}^{90}$Y, we obtain $1-\Gamma_\text{tree}/\Gamma_\text{tot}\approx -7.7\%$, indicating a somewhat smaller correction to the total rate compared to the decay shape; this is due to a partial cancellation between the orange and green lines in Fig.\ref{fig:SrtoYfixEe}, which is purely accidental. One can also study the correction to the beta spectrum, as we present in the supplementary material. Notice that the $\asv q$-integrated results should be taken with a grain of salt as our approximate formula becomes inaccurate in the high-$\asv q$ region. We defer the more comprehensive analysis at arbitrary $\asv q$ to a later work. 

In conclusion, we have  shown  that the existing understanding of decay kinematics in forbidden nuclear transitions has to be thoroughly revisited; the thought-to-be forbidden region of $\asv q\approx 0$ is opened up by RC, and depending on the specific transition the RC contribution may even be larger than the tree-level in a wider kinematic region. On the one hand, this imposes an extra challenge to the theory community due to the need to compute $\asv{k}$-dependent nuclear matrix elements, for instance $\mathfrak C_\text{tree}(\asv{k})$, using reliable \textit{ab initio} methods in order to correctly interpret forbidden beta decay data. On the other hand, our work also unveils a number of new experimental opportunities and discovery potential. By focusing on the small-$\asv q$ region, one effectively evades the large tree-level uncertainty and has a direct experimental probe of the RC physics. 
It is also interesting to notice that, the topologies in Fig.\ref{fig:RC} that open up the forbidden decay at $\asv q=0$ are not only achievable within the SM, but also with new physics. 
While modifications to the charged-current interactions induced by heavy new physics
do not work~\cite{Glick-Magid:2016rsv},
light new DOFs can play a similar role as the SM photon and open up the decay at $\asv q=0$.
Therefore, forbidden decays at small $\asv q$ provide a perfect avenue to study such light new DOFs, provided that the SM RC in this region is computed  to a moderate accuracy; this is of particular interest due to recent observations such as the ATOMKI anomaly~\cite{Krasznahorkay:2015iga}. 
Moreover, given their \%-level size,  the corrections identified in this work must also be included in the analysis of the electron spectra that aim to uncover new physics which originates at high-energy. 
We hope these findings provide new motivations for future theoretical and experimental programs in this topic.

\begin{acknowledgments}
We thank Oscar Naviliat-Cuncic, Mikhail Gorchtein, Leendert Hayen, Charlie Rasco and Guy Ron for inspiring discussions. 
C.-Y.S. and A.G.-M. are supported in part by the U.S. Department of Energy (DOE), Office of Science, Office of Nuclear Physics, under award DE-FG02-97ER41014. Additionally, C.-Y.S. receives support from the FRIB Theory Alliance award DE-SC0013617, and A.G.-M. is supported by the DOE Topical Collaboration "Nuclear Theory for New Physics", award No. DE-SC0023663, and the Hebrew University of Jerusalem through the Dalia and Dan Maydan Post-Doctoral Fellowship. V.C. is supported by the U.S. DOE Office of Science, Office of Nuclear Physics, under Grant No. DE-FG02-00ER41132. 

\end{acknowledgments}

\bibliographystyle{apsrev}
\bibliography{ref}

\clearpage

\setcounter{page}{1}
\renewcommand{\thepage}{Supplementary Information -- S\arabic{page}}
\setcounter{table}{0}
\renewcommand{\thetable}{S\,\Roman{table}}
\setcounter{equation}{0}
\renewcommand{\theequation}{S\,\arabic{equation}}
\setcounter{figure}{0}
\renewcommand{\thefigure}{S\,\arabic{figure}}

\section{Supplementary Material}

\subsection{Rotation of $J=2$ spin states}

The multipole expansion formalism of the Fourier-transformed current operators $J^\mu(\pm\sv{k})$ is built in the coordinate system with $\sv{\hat{z}}\equiv\sv{\hat{k}}$. This frame is problematic when applying to the external (spinful) nuclear state, because $\sv{k}$ is an unobserved photon momentum to be integrated over which cannot be used to represent the direction of the observable external nuclear spin; an observable direction has to be chosen for the latter, and a natural option is the direction of the electron momentum $\sv{\hat{p}_e}$. At the same time, it is necessary to express both the current operators and the nuclear states in the same coordinate system (with $\sv{\hat{z}}=\sv{\hat{k}}$) in order to apply the Wigner-Eckart theorem. Therefore, a transformation matrix of the external nuclear spin state along $\sv{\hat{p}_e}$ and $\sv{\hat{k}}$ is needed. In this work we focus on $J=2$ nuclear states:
\begin{equation}
    |2,m(\sv{\hat{p}_e})\rangle =\sum_{m'=-2}^{2}S_{m'm}(\theta)|2,m'(\sv{\hat{k}})\rangle~,
\end{equation}
where $m$ and $m'$ are the magnetic quantum numbers along $\sv{\hat{p}_e}$ and $\sv{\hat{k}}$, respectively. The matrix $S(\theta)$ reads:\begin{widetext}
	\begingroup
\begin{equation}
	S(\theta)=\left(\begin{array}{ccccc}
		\cos^{4}\frac{\theta}{2} & \cos^{2}\frac{\theta}{2}\sin\theta & \sqrt{\frac{3}{8}}\sin^{2}\theta & \sin^{2}\frac{\theta}{2}\sin\theta & \sin^{4}\frac{\theta}{2}\\
		-\cos^{2}\frac{\theta}{2}\sin\theta & \cos^{2}\frac{\theta}{2}(2\cos\theta-1) & \sqrt{\frac{3}{2}}\cos\theta\sin\theta & \frac{1}{2}(\cos\theta-\cos2\theta) & \sin^{2}\frac{\theta}{2}\sin\theta\\
		\sqrt{\frac{3}{8}}\sin^{2}\theta & -\sqrt{\frac{3}{2}}\cos\theta\sin\theta & 1-\frac{3}{2}\sin^{2}\theta & \sqrt{\frac{3}{2}}\cos\theta\sin\theta & \sqrt{\frac{3}{8}}\sin^{2}\theta\\
		-\sin^{2}\frac{\theta}{2}\sin\theta & \frac{1}{2}(\cos\theta-\cos2\theta) & -\sqrt{\frac{3}{2}}\cos\theta\sin\theta & \cos^{2}\frac{\theta}{2}(2\cos\theta-1) & \cos^{2}\frac{\theta}{2}\sin\theta\\
		\sin^{4}\frac{\theta}{2} & -\sin^{2}\frac{\theta}{2}\sin\theta & \sqrt{\frac{3}{8}}\sin^{2}\theta & -\cos^{2}\frac{\theta}{2}\sin\theta & \cos^{4}\frac{\theta}{2}
	\end{array}\right)~,
\end{equation}
\endgroup
\end{widetext}
where $\theta\equiv\cos^{-1}(\sv{\hat{p}_e}\cdot\sv{\hat{k}})$.

\subsection{Reduced matrix elements}

\begin{table}
	\begin{centering}
		\begin{tabular}{|c|c|c|}
			\hline 
			RME & $i(2^{-})\to f(0^{+})$ & $i(0^{+})\to f(2^{-})$\tabularnewline
			\hline 
			\hline 
			$\mathfrak C_{\text{tree}}$ & $\frac{1}{\asv{k}}\langle f(0^{+})||L_{2}^{A}(\asv{k})||i(2^{-})\rangle$ & $\frac{1}{\asv{k}}\langle f(2^{-})||L_{2}^{A}(\asv{k})||i(0^{+})\rangle$\tabularnewline
			\hline 
			$\mathfrak C_{\gamma g}^{i,f}$ & \multicolumn{2}{c|}{$\frac{1}{Z_{i,f}}\langle J_{i,f}m_{i,f}|J^{0}(\asv{k})|J_{i,f}m_{i,f}\rangle$}\tabularnewline
			\hline 
			$\mathfrak C_{\gamma1}^{X}$ & $\frac{1}{\asv{k}}\langle f(0^{+})||C_{1}^{\text{em}}(\asv{k})||f(1_{X}^{-})\rangle$ & $\frac{1}{\asv{k}}\langle f(2^{-})||C_{1}^{\text{em}}(\asv{k})||f(1_{X}^{+})\rangle$\tabularnewline
			\hline 
			$\mathfrak C_{\gamma2}^{X}$ & $\frac{1}{\asv{k}}\langle i(1_{X}^{+})||C_{1}^{\text{em}}(\asv{k})||i(2^{-})\rangle$ & $\frac{1}{\asv{k}}\langle i(1_{X}^{-})||C_{1}^{\text{em}}(\asv{k})||i(0^{+})\rangle$\tabularnewline
			\hline 
			$\mathfrak C_{A1}^{X}$ & $\langle f(1_{X}^{-})||L_{1}^{A}(\asv{k})||i(2^{-})\rangle$ & $\langle f(1_{X}^{+})||L_{1}^{A}(\asv{k})||i(0^{+})\rangle$\tabularnewline
			\hline 
			$\mathfrak C_{A2}^{X}$ & $\langle f(0^{+})||L_{1}^{A}(\asv{k})||i(1_{X}^{+})\rangle$ & $\langle f(2^{-})||L_{1}^{A}(\asv{k})||i(1_{X}^{-})\rangle$\tabularnewline
			\hline 
		\end{tabular}
		\par\end{centering}
	\caption{Definitions of modified reduced matrix elements.\label{tab:RME}}
\end{table}

\begin{table}
\begin{centering}
\begin{tabular}{|c|c|c|}
\hline 
Decay process & Intermediate states & Multipole operators\tabularnewline
\hline 
\hline 
\multirow{2}{*}{$i(2^{-})\rightarrow f(0^{+})$} & $i(2_{g}^{-}),f(0_{g}^{+})$ & $C_{0}^{\text{em}},L_{2}^{A},E_{2}^{A}$\tabularnewline
\cline{2-3} 
 & $i(1_{X}^{+}),f(1_{X}^{-})$ & $C_{1}^{\text{em}},L_{1}^{\text{em}},E_{1}^{\text{em}},L_{1}^{A},E_{1}^{A}$\tabularnewline
\hline 
\multirow{2}{*}{$i(0^{+})\rightarrow f(2^{-})$} & $i(0_{g}^{+}),f(2_{g}^{-})$ & $C_{0}^{\text{em}},L_{2}^{A},E_{2}^{A}$\tabularnewline
\cline{2-3} 
 & $i(1_{X}^{-}),f(1_{X}^{+})$ & $C_{1}^{\text{em}},L_{1}^{\text{em}},E_{1}^{\text{em}},L_{1}^{A},E_{1}^{A}$\tabularnewline
\hline 
\end{tabular}
\par\end{centering}
\caption{\label{tab:leadingO}Intermediate states and multipole operators responsible for the dominant
contributions to $T^{\mu\nu}$}
\end{table}

The reduced matrix element of a generic multipole operator $O_{Jm}(\asv{k})$ is defined through the Wigner-Eckart theorem:
\begin{eqnarray}
	\langle J_{f}m_{f}|O_{Jm}(\asv{k})|J_{i}m_{i}\rangle&=&(-1)^{J_{f}-m_{f}}\left(\begin{array}{ccc}
		J_{f} & J & J_{i}\\
		-m_{f} & m & m_{i}
	\end{array}\right)\nonumber\\
	&&\times \langle J_{f}||O_{J}(\asv{k})||J_{i}\rangle\:\:.\label{eq:WET}
\end{eqnarray}
In our analysis we need multiple operators from the electromagnetic and the axial weak current, which we denote as $O_{Jm}^{\rm em}$ and $O_{Jm}^{A}$, respectively. 
It is also beneficial to scale out the leading $\asv{k}$- (and $Z$)-dependence and define a new set of reduced matrix elements that are non-zero at $\asv{k}=0$; this is done in Table~\ref{tab:RME}. In particular, we have $\mathfrak C_{\gamma g}^{i,f}(0)=1$.

\subsection{Full leading expression of $T^{\mu\nu}$}

In this work we are only interested in the region where $\asv{k}\sim E_i-E_X\sim E_f-E_X\sim p_e$, where $X$ is a nuclear intermediate state. So, it is convenient to define an expansion parameter $\epsilon\sim p_e R$ and expand the current product according to its power. Standard power counting of multipole operators~\cite{Glick-Magid:2021xty} shows that the leading order terms scale as:
\begin{equation}
    \langle J_{\text{em}}\rangle \times\langle J_{W}\rangle\sim\epsilon^{1}~,
\end{equation}
and are contributed by the operators in Table~\ref{tab:leadingO}; effects of other multipole operators are suppressed by extra powers of $\epsilon$. Additionally,  with the mathematical identity
\begin{equation}
    E_J=\sqrt{\frac{J+1}{J}}L_J\left(1+\mathcal{O}(\epsilon^2)\right)
\end{equation}
and the conservation of vector current that relates the matrix element of $E_J^\text{em}$ and $C_J^\text{em}$, the number of independent multipole operators can be further reduced. With these, we display the full expression of $T^{\mu\nu}$ at leading order of the $\epsilon$-expansion:
\begin{widetext}
\begin{eqnarray}
	T^{0j}(m) & = & i\sqrt{4M_{i}M_{f}}\asv{k}\left[\sqrt{4\pi}\mathfrak C_{\text{tree}}\left(\frac{Z_{f}\mathfrak C_{\gamma g}^f}{k_{0}+i\varepsilon}-\frac{Z_{i}\mathfrak C_{\gamma g}^i}{k_{0}-i\varepsilon}\right) +\sqrt{\frac{32}{5}}\pi\sum_{X}\left(\frac{\mathfrak C_{\gamma1}^{X}\mathfrak C_{A1}^{X}}{M_{f}(1_{X}^{\mp})-M_{f}-k_{0}-i\varepsilon}\right.\right.\nonumber\\
	&&\left.\left.+\frac{\mathfrak C_{\gamma2}^{X}\mathfrak C_{A2}^{X}}{M_{i}(1_{X}^{\pm})-M_{i}+k_{0}-i\varepsilon}\right)\right] \left\{ S_{0m}(\theta)(\boldsymbol{\epsilon_{0}}^{*})^{j}\mp\frac{\sqrt{3}}{2}\left(S_{\mp1m}(\theta)(\boldsymbol{\epsilon_{1}}^{*})^{j}+S_{\pm1m}(\theta)(\boldsymbol{\epsilon_{-1}}^{*})^{j}\right)\right\} \nonumber \\
	T^{jl}(m) & = & i\pi\sqrt{\frac{32}{5}}\sqrt{4M_{i}M_{f}}\sum_{X}\left(\frac{(M_{f}(1_{X}^{\mp})-M_{f})\mathfrak C_{\gamma1}^{X}\mathfrak C_{A1}^{X}}{M_{f}(1_{X}^{\mp})-M_{f}-k_{0}-i\varepsilon}+\frac{(M_{i}-M_{i}(1_{X}^{\pm}))\mathfrak C_{\gamma2}^{X}\mathfrak C_{A2}^{X}}{M_{i}(1_{X}^{\pm})-M_{i}+k_{0}-i\varepsilon}\right)\times\nonumber \\
	&  & \left\{ (\boldsymbol{\epsilon_{0}}^{*})^{j}\left[S_{0m}(\theta)(\boldsymbol{\epsilon_{0}}^{*})^{l}\mp\frac{\sqrt{3}}{2}S_{\mp1m}(\theta)(\boldsymbol{\epsilon_{1}}^{*})^{l}\mp\frac{\sqrt{3}}{2}S_{\pm1m}(\theta)(\boldsymbol{\epsilon_{-1}}^{*})^{l}\right]\right.\nonumber \\
	&  & +(\boldsymbol{\epsilon_{1}}^{*})^{j}\left[\mp\frac{\sqrt{3}}{2}S_{\mp1m}(\theta)(\boldsymbol{\epsilon_{0}}^{*})^{l}+\sqrt{\frac{3}{2}}S_{\mp2m}(\theta)(\boldsymbol{\epsilon_{1}}^{*})^{l}+\frac{1}{2}S_{0m}(\theta)(\boldsymbol{\epsilon_{-1}}^{*})^{l}\right]\nonumber \\
	&  & \left.+(\boldsymbol{\epsilon_{-1}}^{*})^{j}\left[\mp\frac{\sqrt{3}}{2}S_{\pm1m}(\theta)(\boldsymbol{\epsilon_{0}}^{*})^{l}+\frac{1}{2}S_{0m}(\theta)(\boldsymbol{\epsilon_{1}}^{*})^{l}+\sqrt{\frac{3}{2}}S_{\pm2m}(\theta)(\boldsymbol{\epsilon_{-1}}^{*})^{l}\right]\right\} 
\end{eqnarray}
\end{widetext}

\subsection{Correction to the beta spectrum}

\begin{figure}[htb]
	\centering
	\includegraphics[width=1.0\columnwidth]{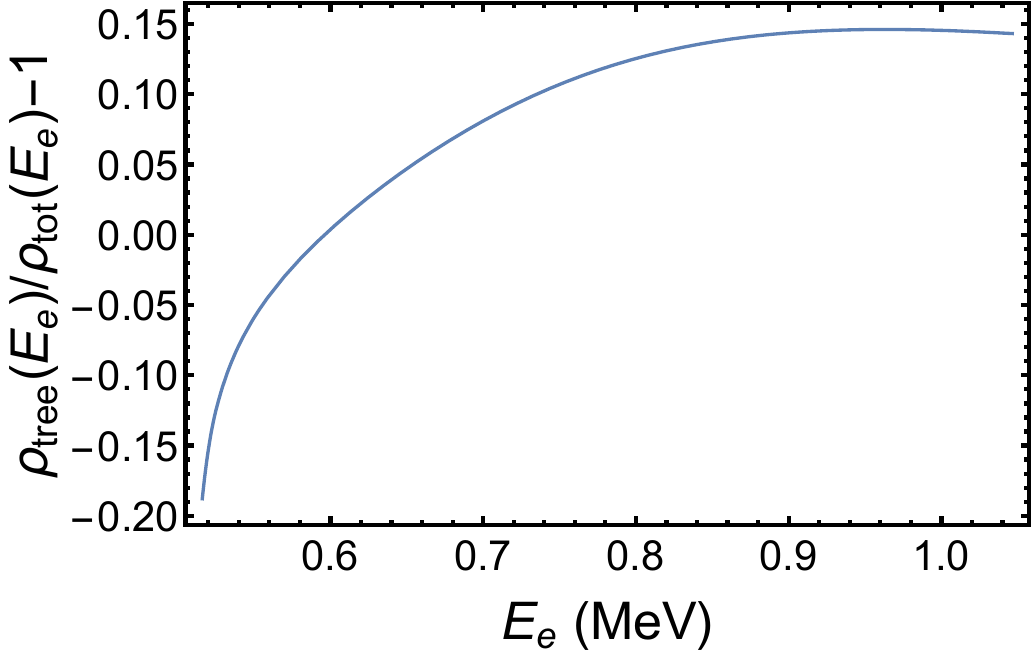}
	\caption{\label{fig:specratio}The correction to the beta spectrum for $^{90}$Sr$\to$$^{90}$Y.} 
\end{figure}

By performing the $q$-integral in Eq.\eqref{eq:totalrate} and leaving $E_e$ unintegrated, we can also study the RC to the beta spectrum, which we plot in Fig.\ref{fig:specratio} for $^{90}$Sr$\to$$^{90}$Y. 

\end{document}